\documentclass[prb,twocolumn,showpacs,superscriptaddress,floatfix]{revtex4-1}
\usepackage{graphicx}
\usepackage{amsmath}
\usepackage{amsfonts}
\usepackage{amssymb}
\usepackage{txfonts}

\newcommand{\beq}{\begin{equation}}
\newcommand{\eeq}{\end{equation}}
\newcommand{\beqa}{\begin{eqnarray}}
\newcommand{\eeqa}{\end{eqnarray}}

\newcommand{\rr}{{\bf r}}

\newcommand{\p}{{\bf p}}
\newcommand{\q}{{\bf q}}

\newcommand{\kk}{{\bf k}}
\newcommand{\om}{\omega}

\newcommand{\gb}{\bar{G}}

\begin{document}

\title{Superfluid transition in disordered dipolar Fermi gases}

\author{S.\,I. Matveenko}
\affiliation{Russian Quantum Center, Skolkovo, Moscow 143025, Russia}
\affiliation{Landau Institute for Theoretical Physics, RAS, 142432, Chernogolovka, Moscow region, Russia}

\author{ V.\,I. Yudson}
\affiliation{Laboratory for Condensed Matter Physics, National Research University
Higher School of Economics, Moscow, 101000, Russia}
\affiliation{Russian Quantum Center, Skolkovo, Moscow 143025, Russia}

\author{B.\,L. Altshuler}
\affiliation{Physics Department, Columbia University, 538 West 120th Street, New York, New York 10027, USA}
\affiliation{Russian Quantum Center, Skolkovo, Moscow 143025, Russia}

\author{G.\,V. Shlyapnikov}
\affiliation{Russian Quantum Center, Skolkovo, Moscow 143025, Russia}
\affiliation{Universit\'e Paris-Saclay, CNRS, LPTMS, 91405 Orsay, France}
\affiliation{Van der Waals-Zeeman Institute, Institute of Physics, University of Amsterdam, Science Park 904, 1098 XH Amsterdam, The Netherlands}

\date{\today}

\begin{abstract}
We consider a weakly interacting two-component Fermi gas of dipolar particles (magnetic atoms or polar molecules) in the two-dimensional geometry. The dipole-dipole interaction
 (together with the short-range interaction at Feshbach resonances) for dipoles perpendicular to the plane of translational motion may provide a superfluid transition. The dipole-dipole
 scattering amplitude is momentum dependent, which violates the Anderson theorem claiming the independence of the transition temperature on the presence of weak disorder.
 We have shown that the disorder can strongly increase the critical temperature (up to $10$ nK at realistic densities). This opens wide possibilities for the studies of the superfluid
 regime in weakly interacting Fermi gases, which was not observed so far.
\end{abstract}
\pacs{ }
\maketitle
\section{Introduction}
The last decades were marked by profound achievements in the physics of ultracold atomic Fermi gases. The key circumstance was the use of Feshbach resonances (magnetic
field dependence of the interaction amplitude) allowing one to change the interaction strength in a wide range, even from an infinite repulsion to infinite attraction \cite{chen2010}. Experiments
with two-component Fermi gases have reached the strongly interacting regime and identified a superfluid transition in this regime \cite{dalibard2008,giorgini2008}, which brings in analogies with neutron stars
and superconductors. However, experimental studies did not achieve the weakly interacting Bardeen-Cooper-Schrieffer (BCS) regime: for common  densities $n\lesssim 10^{14}$ cm$^{-3}$ the
superfluid transition temperature $T_c$ would be about a nanokelvin or lower, i.e. beyond experimental reach.

Possibilities to manipulate the superfluid transition temperature, in particular by manipulating the external confining potential, was always at the core of the studies \cite{giorgini2008}. In the present
stage, after the observation of Anderson localization in dilute clouds of neutral atoms in disorder \cite{Billy,Roati}, the behavior of disordered ultracold quantum gases became a rapidly growing
domain of research \cite{Sanchez-Palencia,ReviewHuse,ReviewAbanin}. One of the  key questions is how the superfluid transition temperature of a two-component Fermi gas can be modified by introducing a disorder.
This question has been the subject of a number of works
in condensed matter  and in cold atomic gases \cite{Anderson,ag,lee,sadovskii,ov,fin,M&F,T&K,Stri}.
As was pointed out by several authors \cite{Dis-at-AT} the critical temperature can be increased when approaching the Anderson transition.
%
%
In a weak disorder ($k_Fl\gg 1$,
where $k_F$ is the Fermi momentum, and $l$ the mean free path) and for the case of weak short-range interaction
 where the  interaction amplitude is momentum independent,  one has the Anderson theorem \cite{Anderson}: the BCS
transition temperature is disorder independent. In a later stage, this statement was justified by Abrikosov and Gor'kov \cite{ag} within the diagrammatic approach.
However, the works \cite{Anderson,ag} do not take into account weak localization effects \cite{AAL-DOS}
which, in the presence of interaction, change the fermion self-energy and the density of states. Including these corrections the disorder leads to a moderate increase of the BCS transition temperature (in the absence
of Coulomb interactions) \cite{M&F,T&K}.

In this paper we consider a two-component two-dimensional (2D) gas of dipolar fermions (magnetic atoms or polar molecules) in a weak disorder, assuming that the dipoles are perpendicular  to the plane
of the translational motion. This can be a mixture of two different isotopes of magnetic atoms in the lowest Zeeman states (for example, fermionic isotopes of dysprosium which has
magnetic moment of $10\mu_B$, and we will omit a small difference in masses of these isotopes). In this geometry the dipole-dipole interaction amplitude  by itself consists of a fairly large
short-range repulsive contribution \cite{note1} and a long-range attractive momentum-dependent contribution, so that the total amplitude is positive. However, the short-range repulsion (complemented by the
non-dipole contribution) can be strongly reduced or even converted to attraction by using Feshbach resonances. This can make the total interaction amplitude attractive and provide a superfluid transition like
in bilayer dipolar systems \cite{pik2010}. Since the amplitude is now momentum-dependent,  in the presence of weak disorder the Anderson theorem does not work.

Strictly speaking, in two dimensions we have the Kosterlitz-Thouless superfluid phase transition. However, in the weakly interacting regime the transition temperature is very close to that calculated
in the Bardeen-Cooper-Schrieffer (BCS) approach \cite{miyake}.
We find that the momentum dependence of the interaction amplitude by itself may lead to a significant increase of the BCS transition temperature in the presence of disorder. The weak localization corrections work
in the same direction. As a result, the BCS transition temperature can be strongly increased by the disorder, which opens wide possibilities for the observation of superfluidity in weakly
interacting Fermi gases of magnetic atoms and/or polar molecules.

The paper is organized as follows. In section II we present a general formalism for studying the Cooper pairing instability in the presence of disorder. Section III contains our derivation of
the critical temperature $T_c$ omitting weak localization corrections. These corrections are taken into account in section IV, where we present the final result for the increase
of $T_c$ by the disorder. In section V we conclude.

\section{Cooper pairing instability in disordered Fermi systems. General formalism}
\begin{figure}
  \centering
  \includegraphics[width=1.0\linewidth]{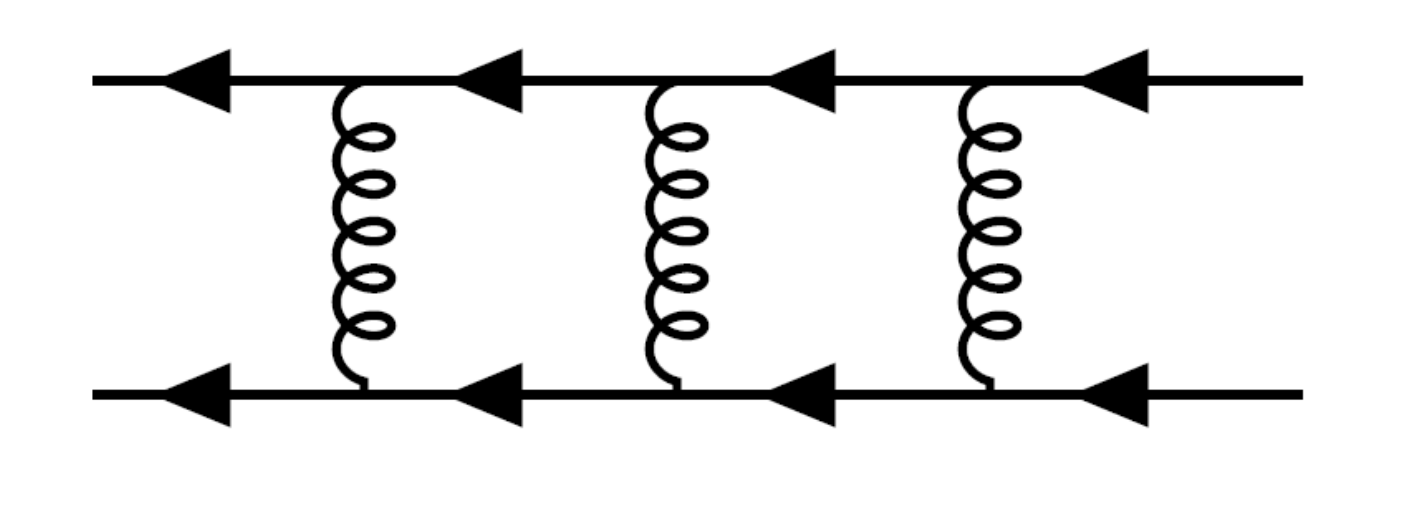}
  \caption{ \label{wl1} Diagrammatic representation for the susceptibility $\chi$. The upper and
lower lines correspond to fermionic Green functions and the wavy lines
correspond to the interaction potential $V(r_1 , r_2)$.
}
\end{figure}

The threshold of the Cooper pairing instability in a system of weakly interacting two-component fermions is determined by a singularity that occurs at a critical temperature $T_c$ in the
susceptibility function $\chi(\mathbf{r}, \mathbf{\bar{r}}; \mathbf{r'}, \mathbf{\bar{r'}})$, which describes the system response $\langle \psi_{\uparrow}(\mathbf{r})\psi_{\downarrow}(\mathbf{\bar{r}})\rangle$
to a perturbation of the form $\int d\mathbf{r'} d\mathbf{\bar{r'}} \psi^{\dag}_{\uparrow}(\mathbf{r'}) \psi^{\dag}_{\downarrow}(\mathbf{\bar{r'}})h(\mathbf{r'}, \mathbf{\bar{r'}})$. Here $\psi_{\uparrow}(\mathbf{r})$
and $\psi_{\downarrow}(\mathbf{r})$ are annihilation operators of fermionic components, let say spin up and spin down. For weakly interacting fermions the diagrammatic representation of $\chi$
corresponds to a series of ladder diagrams, where the upper and lower fermionic lines are connected by non-intersecting (wavy) lines associated with the interaction potential $V(\mathbf{r}_1 - \mathbf{r}_2)$,
see Fig.~\ref{wl1}. Symbolically, the ladder series corresponds to an infinite sum

\beq
\tilde{B} + \tilde{B}(-V)\tilde{B} + \tilde{B}(-V)\tilde{B}(-V)\tilde{B}... = \tilde{B}[I + V\tilde{B}]^{-1},
 \label{ladder}
 \eeq
where
\beq
\tilde{B} = T_c\sum_{\varepsilon_n}B(\varepsilon_n) ,
 \label{tilde-B}
 \eeq
and $B(\varepsilon_n)$ is an elementary block of two fermionic Green functions: $B(\varepsilon_n)=G(\varepsilon_n)G(-\varepsilon_n)$. The summation in Eq.(\ref{tilde-B}) runs over the fermion
Matsubara frequencies $\varepsilon_n = 2\pi T_c(n+1/2)$, $n= 0, \pm 1, ...$. In a clean system (without any disorder), the instability of the ladder series corresponds to a zero eigenvalue
of a linear integral operator $I+V\tilde{B}$ or, in other words, to the existence of a non-zero eigenfunction $\Delta$ obeying the (symbolic) equation
\beq
 \Delta = -V\tilde{B}\Delta ,
 \label{Delta}
 \eeq
which is the standard equation for $T_c$.

In the presence of disorder, one should associate the instability threshold
with the singularity of the susceptibility function
$\chi$ \emph{averaged over the disorder}. In the case of a weak disorder
(such that $k_F l \gg 1$), the operator $\tilde{B}$ in the
equation for $\Delta$ is replaced with a new operator $\tilde{\mathcal{B}}$.
There are two kinds of modifications.
First, the elementary block $B(\varepsilon)$ is replaced with its disorder
averaged value
\beq
B_{av}(\varepsilon_n)= <G(\varepsilon_n)G(-\varepsilon_n)> ,
 \label{B-av}
 \eeq
After the summation over the Matsubara frequencies this block gives a contribution $\tilde{B}_{av}= T_c\sum_{\varepsilon_n}B(\varepsilon_n)$
to the integral operator
\beq
\tilde{\mathcal{B}}= \tilde{B}_{av} +\delta \tilde{\mathcal{B}}
 \label{B-cal}
 \eeq
of the disordered system. The second contribution
\beq
\delta \tilde{\mathcal{B}} =  \delta \tilde{B}_{\Sigma} + \delta \tilde{B}_{V},
 \label{delta-B}
 \eeq
originates not from averaging the elementary block but from the disorder-induced
corrections $\delta \Sigma$ to the fermion self-energy and to the fermion interaction
$V$ (the so called vertex corrections). These ''weak localization'' (WL)
corrections were studied quite some time ago. Corrections to the self-energy and
the corresponding WL corrections to the density of states were considered
in the pioneer paper \cite{AAL-DOS}. The influence of WL corrections on the critical temperature
of superconducting transition was explored in \cite{M&F} and in the later work \cite{T&K}.
We shall discuss the significance of these corrections later. The relative smallness of
the disorder-induced corrections allows one to calculate them independently of each other.

In the present section and in the next one we are returning to the study of the first kind of corrections to $T_c$, which are caused by
the contribution $\tilde{B}_{av}$ (\ref{B-av}) to the kernel $\tilde{\mathcal{B}}$.
These corrections are sensitive to the particular spatial dependence of the interaction potential.
For instance, they are absent for the contact interaction (in accordance with the Anderson theorem \cite{Anderson}).
On the contrary, we will show that for the dipole-dipole interaction that we are interested in the corrections
are nonzero and can dominate over the WL corrections. Details of the calculation are presented in the
next section. Here we only describe the structure of the averaged block $B_{av}(\varepsilon_n)$.
As is well known, the leading correction to the averaged (over a weak disorder) product of two
Green functions with opposite frequencies and incident wave vectors is given by the ladder of
parallel impurity lines connecting two fermionic lines. Such a ''Cooperon" installation
bears the total zero momentum and depends on the difference between the two frequencies, $\varepsilon_n$
and $-\varepsilon_n$, i.e., on $2\varepsilon_n$. Thus, the averaged block $B_{av}(\varepsilon_n)$ has the form
\beq
B_{av}(\varepsilon_n) = B_0(\varepsilon_n) + B_0(\varepsilon_n)\Gamma(\varepsilon_n)B_0(\varepsilon_n),
 \label{averaged-Block}
 \eeq
where $B_0(\varepsilon_n) = \langle G(\varepsilon_n)\rangle \langle G(-\varepsilon_n)\rangle$
is the product of two disorder-averaged Green functions (a more detailed definition is given below in Eq.(\ref{GG})), and the quantity
\beq
\Gamma(\varepsilon_n)= \gamma\frac{1 + 2\tau |\varepsilon_n |}{2\tau |\varepsilon_n |},
 \label{Gamma}
 \eeq
results from the Cooperon carrying zero total momentum \cite{Note}. The parameter $\gamma$ comes from the correlation
function for a short range disorder potential $U({\bf r})$, namely $<U({\bf r})U({\bf r}')>=\gamma\delta({\bf r}-{\bf r}')$.
The time $\tau$ is the inverse disorder-induced scattering rate $1/\tau=2\pi \rho_F\gamma$,  and
$\rho_F$ is the density of states on the Fermi surface.

\section{Derivation without WL corrections}
Taking into account only the averaged block  $ \tilde{B}_{av}$  in Eq.(\ref{B-cal})  we rewrite the
equation $\Delta = -V \tilde{B} \Delta$    at $T \to  T_c $  in the form
 \beq
\begin{split}
 \Delta (\rr -\rr') = - V(\rr - \rr') T \sum_{\om_n} \int d \rr_1 d\rr_2  \\
\langle G(\rr, \rr_1 ; \om_n) G(\rr_2, \rr';-\om_n)\rangle \Delta(\rr_1 - \rr_2),
 \end{split}
\label{}
 \eeq
 or in the momentum representation
  \beq
\begin{split}
 \Delta (\kk) = -\sum_{\p, \kk'} V(\kk - \p) T \sum_{\om_n}  \\
 \langle G(\p, \kk' ; \om_n)
  G(\kk', \p;-\om_n)\rangle \Delta(\kk'),
\end{split}
\label{}
 \eeq
where the normalization volume is put equal to unity.
After averaging over the disorder and using Eqs. (\ref{averaged-Block}) and (\ref{Gamma}) we obtain
\begin{widetext}
 \beq
 \Delta (\kk) = -\sum_{\p, \kk'} V(\kk - \p) T \sum_{\om_n}  \left[ \left[\delta_{\p, \kk'}
 + \gb (\p,  \om_n) \gb( \p, -\om_n) \frac{\gamma (1+2|\om_n| \tau)}{2 |\om_n| \tau}\right]
 \gb(\kk',\om_n) \gb(\kk',-\om_n)\right] \Delta(\kk'),
 \label{full}
 \eeq
\end{widetext}
where the averaged  Green function is
 \beq
 \bar{G}(\kk, \om_n) = \frac{1}{i \om_n +\frac{i}{2\tau} \mbox{sgn}\,\om_n - \xi_{\kk}},
 \label{}
 \eeq
 with $\xi_{\kk} = \frac{k^2}{2m} - \mu$, and $\mu$ the chemical potential (hereinafter $\hbar=1$).

 We  now represent $\Delta({\bf k}')$ in the rhs of Eq.(\ref{full}) as
 $\Delta(\p) + [\Delta({\bf k}' ) - \Delta(\p )]$ and argue later that the second term gives a  small contributions  and can be neglected.

 Then we  make a summation over  $k'$  by using the identity
 \beq
B_0(\varepsilon_n) = \sum_{\q}  \gb (\q,  \om_n) \gb( \q, -\om_n)= \frac{\gamma^{-1}}{1 + 2 |\om_n|\tau},
  \label{GG}
  \eeq
and arrive at the equation:
 \beq
 \Delta(\kk) = - 2T_c \sum_{n\geq 0} \int \frac{d^2 \p}{(2 \pi)^2} \frac{V(\kk - \p)\Delta({\p})\left(1+\frac{1}{2 \om_n \tau}\right)}{\xi^2 +\left(\om_n +\frac{1}{2 \tau}\right)^2},
 \label{short}
\eeq
with $\om_n = \pi T_c (2 n +1)$.  After  the summation over the frequencies we obtain:
\beq
\Delta(\kk) = -     \int \frac{d^2 \p}{(2 \pi)^2}V(\kk - \p) \Delta(\p) K(\p),
\label{bcs}
\eeq
where
\beq
K(\p) = \frac{i}{2\pi} \frac{ \Psi\left(\frac{1}{2} - \frac{i z_{\p}}{2\pi T_c}\right)
 -\Psi(\frac{1}{2})}{z_{\p}} + c.c.,
 \label{Kp}
 \eeq
 $\Psi(x) \equiv\Gamma'(x)/\Gamma(x) $ the digamma function, $z_{\p} = \xi_{\p} + \frac{i}{2\tau}$,  and near the Fermi surface one has $\xi_{\p} \approx v_F (p - p_F)$ with $v_F$ being the Fermi velocity. We then have
 \[ {\mbox Im} \Psi \left(\frac{1}{2} + \frac{i \xi}{2 \pi T_c}\right)=\frac{\pi}{2} \tanh \frac{\xi}{2 T_c},
 \]
\begin{widetext}
\beq
K(p) =\frac{1}{\pi}\frac{1}{\xi_p^2 +\frac{1}{(2 \tau)^2}}\left[ \frac{1}{2\tau}{\mbox Re}\left[\Psi\left(\frac{1}{2} +\frac{i\xi_p}{2 \pi T_c}+\frac{1}{4 \pi \tau T_c}\right) - \Psi \left(\frac{1}{2}\right)\right] +\xi_p\, {\mbox Im}\left[\Psi \left(\frac{1}{2} +\frac{i \xi_p}{2 \pi T_c} + \frac{1}{4 \pi \tau T_c}   \right) \right]\right].
\label{}
\eeq
\end{widetext}
In the limit of $1/\tau \to 0$ we have
\beq
K(p) \to \frac{\tanh\frac{\xi_p}{2T_c}}{2\xi_p} \equiv K_0(p)
\label{}
\eeq
Note  that  for the contact potential  (momentum independent)    $V(p) = const $  the transition temperature
is independent of the disorder (Anderson theorem). To see this we should shift the integration contour as $\xi \to \xi - i/2\tau$
in Eq.(\ref{bcs}) and use analytical properties of the digamma function.  As a result we get the clean case equation.

Using the relation between the potential $V({\bf k}'-{\bf k})$ and the off-shell scattering amplitude $f(\kk', \kk)$:
\beq
f(\kk', \kk)= V((\kk'-\kk) + \int \frac{d^2\q}{(2\pi)^2} \frac{V(\kk'-\q) f(\q, \kk)}{2(E_k -E_q - i0)},
\label{}
\eeq
we find (see, e.g. \cite{levinsen2011})
\beq
\Delta(\kk) = - \fint \frac{d^2\kk'}{(2\pi)^2} f(\kk', \kk) \Delta(\kk') \left[K(\kk') -\frac{1}{2(E_{\kk'} -E_{\kk})}\right].
\label{Tcv}
\eeq
Expanding  the order parameter $\Delta(\kk)$ and the scattering amplitude in a series over the states with different orbital quantum numbers:
$\Delta (\kk ) = \sum_{m=-\infty}^{\infty} \Delta_m(k) \exp (i m \phi_k)$;  $ f(\kk', \kk) = \sum_{m=-\infty}^{\infty} f_m (k', k) \exp [ i m ( \phi_{k'} -\phi_k)]$, we focus on the s-wave symmetry ($m=0$) of the order parameter
and (omitting index $m=0$) obtain from Eq.(\ref{Tcv}):

\beq
\Delta(k) = -\fint \frac{k dk} {2\pi} f(k', k) \Delta(k') \left[K(k') -\frac{1}{2(E_{k'} -E_{k})}\right],
\label{Tc}
\eeq
with  the amplitude $f(k' , k)$ given below.

The scattering amplitude contains two terms,  due to the local  and nonlocal (dipole-dipole) interactions.
For the s-wave scattering the nonlocal part  is given by the integral
\beq
\begin{split}
\int \frac{d^2_0}{r^3} (J_0(k' r)J_0(k r) -1) 2 \pi r \,dr \\
=2 \pi d^2 \left\{
\begin{array}{rcl}
-k F(-1/2, -1/2, 1, k'^2/k^2), &k' < k &\\
-k' F(-1/2, -1/2, 1, k^2/k'^2, & k <k' &.\\
\end{array}
\right.
\end{split}
\label{}
\eeq
Since the hypergeometric function slowly varies in the  interval (0,1):    $ 1< F(...,x) < 4/\pi$,
we put approximately $F (..) =4/\pi $, which is the value on the Fermi surface,  so that
\beq
f(k', k) = F_0 - 8 d^2 {\mbox max}(k, k'),
\label{fkk'}
\eeq
and $f(k_F, k_F)\equiv f_0= F_0 -  8 d^2 k_F <0$. The local part $F_0$ is momentum independent \cite{note1}
and can be varied by the use of Feshbach resonances.

To find the critical temperature we use the ansatz for the order parameter (see \cite{levinsen2011}),
which follows from Eq. (\ref{Tc}) assuming that the main contribution to the integral comes from $k'$ close to $k_F$:
\beq
\Delta(k) = \Delta(k_F) \frac{f(k_F, k)}{f(k_F, k_F)}.
\label{}
\eeq
For $k =k_F$   Eq.(\ref{Tc}) takes the form
\beq
1= - \fint  \frac{d^2\kk'}{(2\pi)^2}   \frac{(f(k', k_F))^2}{f(k_F, k_F)}\left[K(k') -\frac{1}{2(E_{k'} - E_F)}\right],
\label{Tcc}
\eeq
Near the Fermi surface we have
\beq
E_k - E_F \equiv \xi_k \approx v_F (k - k_F); \quad \int  \frac{d^2 \kk}{(2\pi)^2} \approx \int m\frac{d\xi}{2\pi},
\label{nearTc}
\eeq
and
\beq
f(k', k_F) = f_0 - \frac{8 d^2}{v_F} \xi_{k'} \theta (k' - k_F).
\label{fnTc}
\eeq
After the integration in Eq.(\ref{Tcc}) we obtain the  equation

\beqa
& \lambda\ln \frac{T_c}{ T^0_c} = \frac{1}{k_F l} \frac{2 k_F r_*}{\pi^2}\left[ \ln^2\frac{\mu}{2\pi T_c} -2\Psi\left(\frac{1}{2}\right) \ln\frac{\mu}{2\pi T_c} \right] \nonumber \\
&+\frac{1}{k_F l} \frac{4 (k_F r_*)^2}{\pi^3 \lambda}\left[2\ln\frac{\mu}{2\pi T_c} -2-2\Psi\left(\frac{1}{2}\right)\right] \nonumber\\
&-\frac{1}{(k_Fl)^2}  \frac{(k_F r_*)^2}{\pi^2 \lambda}  \ln  \frac{2 \mu e^{C}}{\pi T_c},
\label{eqnTc}
\eeqa
where the mean free path is  $l = v_F \tau$, $C = -\Psi(1)=0.577$, and $r_* = m d^2$ is the dipole-dipole distance. The quantity $T_c^0 = \frac{2 \mu e^{C}}{\pi} \exp (-1/\lambda)$ is the critical temperature in the absence of disorder,  and $\lambda=|f_0|m/2\pi\ll 1$. Detailed calculations leading to Eq.(\ref{eqnTc}) are given in the Appendix.

The terms in the rhs of Eq.(\ref{eqnTc}) should be small (strictly speaking, much smaller than unity). It is this condition that allows us to omit higher order disorder corrections, i.e. terms that are higher order in $1/k_Fl$. In the BCS regime one has $\ln(\mu/T_c)\sim 1/\lambda\gg 1$ and, hence, the second term in the first line of the rhs of Eq.(\ref{eqnTc}) can be omitted. As we consider
the case where $k_F r^* \ll 1$ and  $k_F l \gg 1$,  the last two terms in the rhs of Eq.(\ref{eqnTc}) contain additional small parameters $ k_F r_*$ and $k_F r_* /k_Fl$, and can also be neglected. Thus, equation (\ref{eqnTc}) reduces to
\beq
\ln\frac{T_c}{T^0_c} \approx \frac{2r_*}{\pi^2l}\frac{1}{\lambda^3},
\label{newTc}
\eeq
and the rhs of Eq.(\ref{newTc}) should be significantly smaller than $1/\lambda$. For $r_*/l  = 0.2$, decreasing $\lambda$ from $0.2$ to $0.15$ we obtain $T_c/T_c^0$ increasing from $1.3$ to $1.8$.
Importantly, comparing the result of Eq.(\ref{newTc}) with that of original equation (\ref{eqnTc}) we see that the former is valid within a few percent of accuracy.
Note that we used the simplified equation (\ref{short}) instead of Eq.(\ref{full}).
 A simple but cumbersome  calculation shows  that   omitted  terms  give only  a small contribution to the third
 line in the rhs of Eq. (\ref{eqnTc}).

\section{Influence of weak localization corrections on the disorder-induced increase of $T_c$}
\begin{figure}
  \centering
   \includegraphics[width=3in]{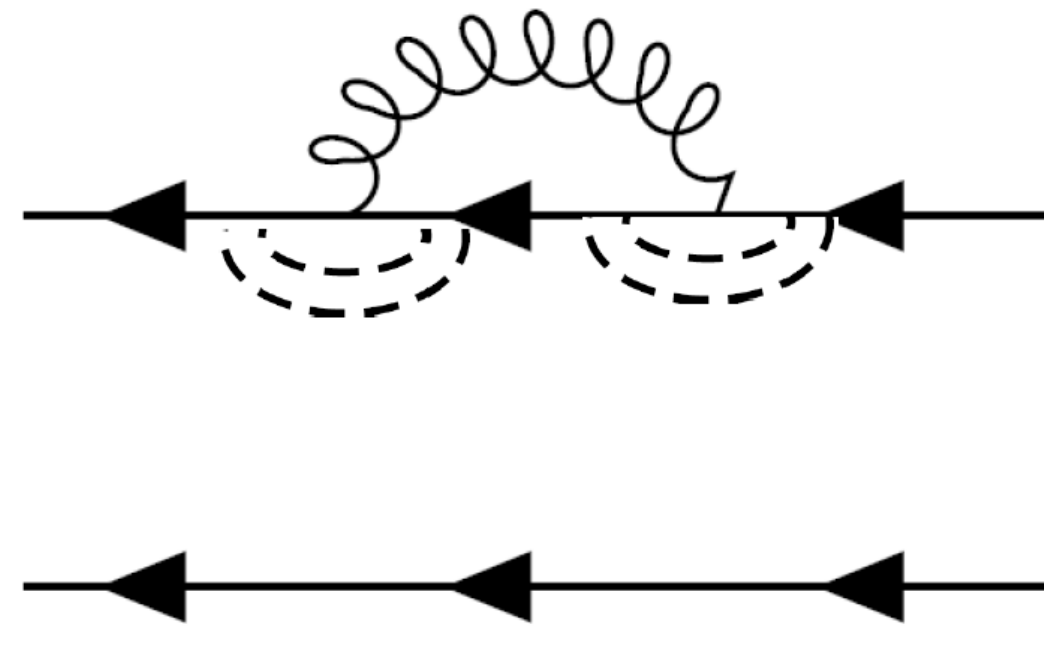}
  \caption{ \label{Fig-WL-corr1} An example of a diagram contributing to the self-energy WL correction.
Here the double dashed lines correspond to the diffuson (ladder) propagator.}
\end{figure}

\begin{figure}
  \centering
   \includegraphics[width=1.0\linewidth]{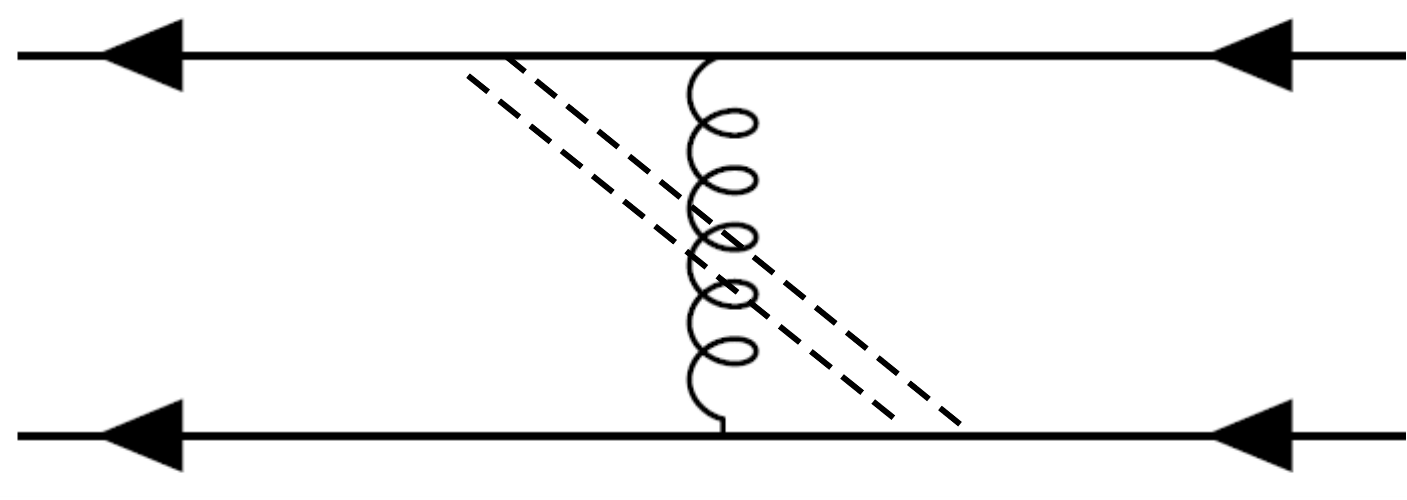}
  \caption{\label{Fig-WL-corr2} An example of a diagram contributing to the vertex WL correction.
Here the double dashed lines correspond to the Cooperon (ladder) propagator.}

\end{figure}

The WL corrections for the disorder-induced change of the critical temperature $T_c$ have been calculated in Refs. \cite{M&F,T&K}.
The WL corrections by themselves lead to the following ratio of $T_c$ to the
critical temperature $T_{c0}$ in the system without disorder:
\beq
\begin{split}
\ln{\left(\frac{T_c}{T^0_c}\right)_{WL}} =
\frac{(3g_F - g_0)\rho_F}{4\pi E_F\tau}\ln^2{\left(\frac{1}{\tau T_c}\right)} \\
- \frac{(g_F + g_0)\rho_F}{6\pi E_F\tau}\ln^3{\left(\frac{1}{\tau T_c}\right)}.
\end{split}
\label{WL}
\eeq
The quantity $g_F$ is defined as $g_F \equiv \overline{V(\mathbf{k} - \mathbf{k'})}$,
where the bar means the angular average of the interaction potential (in the momentum
representation) on the Fermi surface, $k=k'=k_F$. The quantity $g_0$ is
$V(\mathbf{q}=0)$, i.e., the interaction potential with zero momentum transfer.
The first term in Eq.(\ref{WL}) results from the self-energy WL corrections, whereas
the second one originates from the vertex WL corrections. Some of the corresponding
diagrams are shown in Fig.~\ref{Fig-WL-corr1},~\ref{Fig-WL-corr2}. The paired dashed lines there resemble
schematically the ladder diagrams connected by the disorder lines (so called diffuson
and Cooperon diagrams). Equation (\ref{WL}) has been derived under
the assumption $\tau T_c \ll 1$, where the Cooperon and diffusons are large in the
low momentum and low energy limit. The condition $\tau T_c \ll 1$  means that the mean
free path $l= v_F\tau$ is small compared to the correlation length $v_F/T_c$, i.e.,
the motion has a diffusive character. In this diffusive regime
$\ln{\left(\frac{1}{\tau T_c}\right)} \gg 1$, so that the second term in Eq.(\ref{WL})
should be considered as the leading one.

We first express the quantities  $g_F$ and $g_0$ in terms of the scattering amplitude
by using the relation between the potential $V(\mathbf{k'}-\mathbf{k})$ and the off-shell
scattering amplitude $f(\mathbf{k'}, \mathbf{k})$. In the lowest order
(appropriate for the discussed corrections) one has $V(\mathbf{k'}- \mathbf{k})
\approx f(\mathbf{k'}, \mathbf{k})$. Hence, the quantity $g_F$ coincides with
the on-shell amplitude of the $l=0$ channel, i.e., $g_F = f_{l=0}(k_F, k_F)$.
This amplitude, denoted as $f_0$, is given by Eq.(\ref{fkk'}) with $k=k'=k_F$ in the previous section.
Hence, we have
\beq
g_F \rho_F = \frac{f_{0}m}{2\pi}= -\lambda.
\label{gF-nu}
\eeq
The quantity $g_0$ coincides with the on-shell amplitude of the forward scattering:
$g_0= f(\mathbf{k}, \mathbf{k})$ with $k=k_F$. This amplitude is represented as
a sum of partial scattering amplitudes with all momenta $l$:
\beq
f(\mathbf{k}, \mathbf{k}) = f_{l=0}(k, k) + \sum_{l \neq 0} f_{l}(k, k)
\label{}
\eeq
The scattering amplitudes with $l\neq 0$ can be omitted for the short-range interaction, and for the dipole-dipole tail in 2D
they have been calculated in \cite{Lu&GS-2012}. In the limit
$k_Fr_{*}  \ll 1$, the leading contribution to these
partial amplitudes comes from large distances where the interaction can
be treated in the first Born approximation:
\beq
f_{l\neq 0}(\mathbf{k}; \mathbf{k}) \approx \frac{8k r_{*}}{m}\frac{1}{4l^2-1}.
\label{fln0}
\eeq
Making a summation over $l$ in Eq.(\ref{fln0}) we arrive at the expression for the quantity $g_0$:
\beq
g_0 =  f_{0}(k_F) + \frac{8k_Fr_{*}}{m} =
f_{0}(k_F) + 8k_F d^2
\label{g0}
,
\eeq
and, respectively,
\beq
g_0\rho_F = -\lambda + \frac{4 k_Fr_{*}}{\pi}
\label{g0-nu}
.
\eeq

Relative smallness of the WL corrections allows one to replace
$T_c$ by $T^0_c$ in the rhs of Eq.(\ref{WL}) and to represent the latter
in the form
\beq
\ln{\left(\frac{T_c}{T^0_c}\right)}_{WL} \approx
\frac{2\lambda - 4k_Fr_{*}/\pi}{3\pi k_F l}\ln^3{\left(\frac{1}{\tau T^0_c}\right)}
\label{WL-short}
,
\eeq
where we have kept only the leading term with the third power of the large logarithm.
Equation (\ref{WL-short}) originates from the vertex corrections
and can be interpreted as a renormalization of the coupling constant:
$\lambda \rightarrow \lambda + \delta\lambda$, where $\frac{\delta\lambda}{\lambda} \sim \frac{1}{k_Fl}\ln{\left(\frac{1}{\tau T^0_c}\right)}$.
To provide the validity of the approach, the relative correction $\delta\lambda/\lambda$
should be small. This requirement results in the condition
\beq
k_Fl \gg \ln{\left(\frac{1}{\tau T^0_c}\right)} = \frac{1}{\lambda} +
\ln{\left(\frac{\pi e^{-\gamma}}{k_F l}\right)} \gg 1.
\label{Log}
\eeq

Making a summation of the correction to $\ln(T_c/T_{c0})$ due to the momentum dependence of the dipole-dipole scattering amplitude and the WL correction we arrive at the final result:
 \beq
\ln{\left(\frac{T_c}{T^0_c}\right)} \approx \frac{2r^*}{\pi^2 l \lambda^3}
+ \frac{2\lambda - 4k_Fr_{*}/\pi}{3\pi k_F l}\left[\frac{1}{\lambda} +
\ln{\left(\frac{\pi e^{-\gamma}}{k_F l}\right)} \right]^3
.
\label{ln(Tc)-answer}
\eeq
The validity of our approach requires
several conditions, including Eq.(\ref{Log}) and $k_Fr_{*} \ll 1$.

For instance, for the choice $\lambda =0.2$, $r_{*}/l = 0.01$, $k_F r_{*} = 0.1$,
and $k_F l = 10$ we obtain an appreciable increase of the critical temperature:
$T_c \approx 1.4 T^0_c$. Moreover, decreasing $\lambda$ to $0.15$ we find $T_c\approx 2.3 T_{c0}$.
In these cases and also for intermediate values of $\lambda$ the correction to $T_c$ caused by the
momentum dependence of the dipole-dipole interaction amplitude exceeds the WL correction.

We thus see that the momentum dependence of the amplitude of long-range dipolar interaction is crucial
for the disorder-induced increase of the critical temperature, and the overall ratio $T_c/T_{c0}$ may exceed factor 2
for realistic parameters.

\section{Conclusions and outlook}
In conclusion, we have shown that the superfluid transition temperature of a weakly interacting two-component dipolar Fermi gas
can be strongly increased by introducing disorder in the system. The origin of this phenomenon lies in the density fluctuations
caused by the disorder.
Our results can be tested in experiments with magnetic atoms and/or polar molecules. Consider a mixture of dysprosium fermionic
isotopes, $^{161}$Dy and $^{163}$Dy, with equal concentrations and in the lowest  Zeeman states. In the 2D geometry obtained by
strongly confining the atoms in one direction, we orient their magnetic moments (equal to $10\mu_B$) perpendicularly to the plane of
translational motion and thus create the system described in the previous sections. For dysprosium atoms we have $r_*\simeq 200$
angstroms and for the 2D density $n=10^9$ cm$^{-2}$ of each of the components the Fermi momentum and energy are
$k_F\simeq 1.1\times 10^5$ cm$^{-1}$ and $E_F=k_F^2/2m\simeq 300$ nK, so that $k_Fr_*\simeq 0.22$. Selecting the disorder such that
the mean free path is $l=6\times 10^{-5}$ cm and, hence, $k_Fl\simeq 7$ and arranging $\lambda=0.25$ with the use of Feshbach resonances,
equation (\ref{ln(Tc)-answer}) yields $T_c\simeq 2T_{c0}$. At densities specified above the superfluid transition temperature is $T_{c0}\simeq 5$ nK
and, accordingly, the critical temperature in the presence of disorder will be close to $10$ nK. These temperatures are realistic for ongoing experiments
with Fermi gases and they have already been achieved \cite{navon2010,mak2014}. In principle, we can decrease $\lambda$ and obtain a significantly larger ratio $T_c/T_{c0}$. However, the absolute
values of the critical temperature will be significantly lower and likely beyond experimental reach.
Future prospects may concern various types of geometries, for example a bilayer system of dipolar fermions. In this case Cooper pairs can be formed by
fermions belonging to different layers and transform to interlayer bosonic dimers with decreasing the interlayer spacing \cite{pik2010}.  The influence of disorder on the
superfluid transition temperature in this case requires a separate analysis.

\section*{Acknowledgements}
We acknowledge fruitful discussions with M.A. Baranov, I.S. Burmistrov, F. Ferlaino,
L. Chomaz, A.V. Turlapov, and T. Pfau. This work was supported by the Russian
Science Foundation Grant No. 20-42-05002Y.

\vspace{0.5cm}
\appendix*

\section{Calculation of the disorder-induced increase of the critical temperature due to momentum dependence of the interaction amplitude}
\begin{figure}
  \centering
   \includegraphics[width=3in]{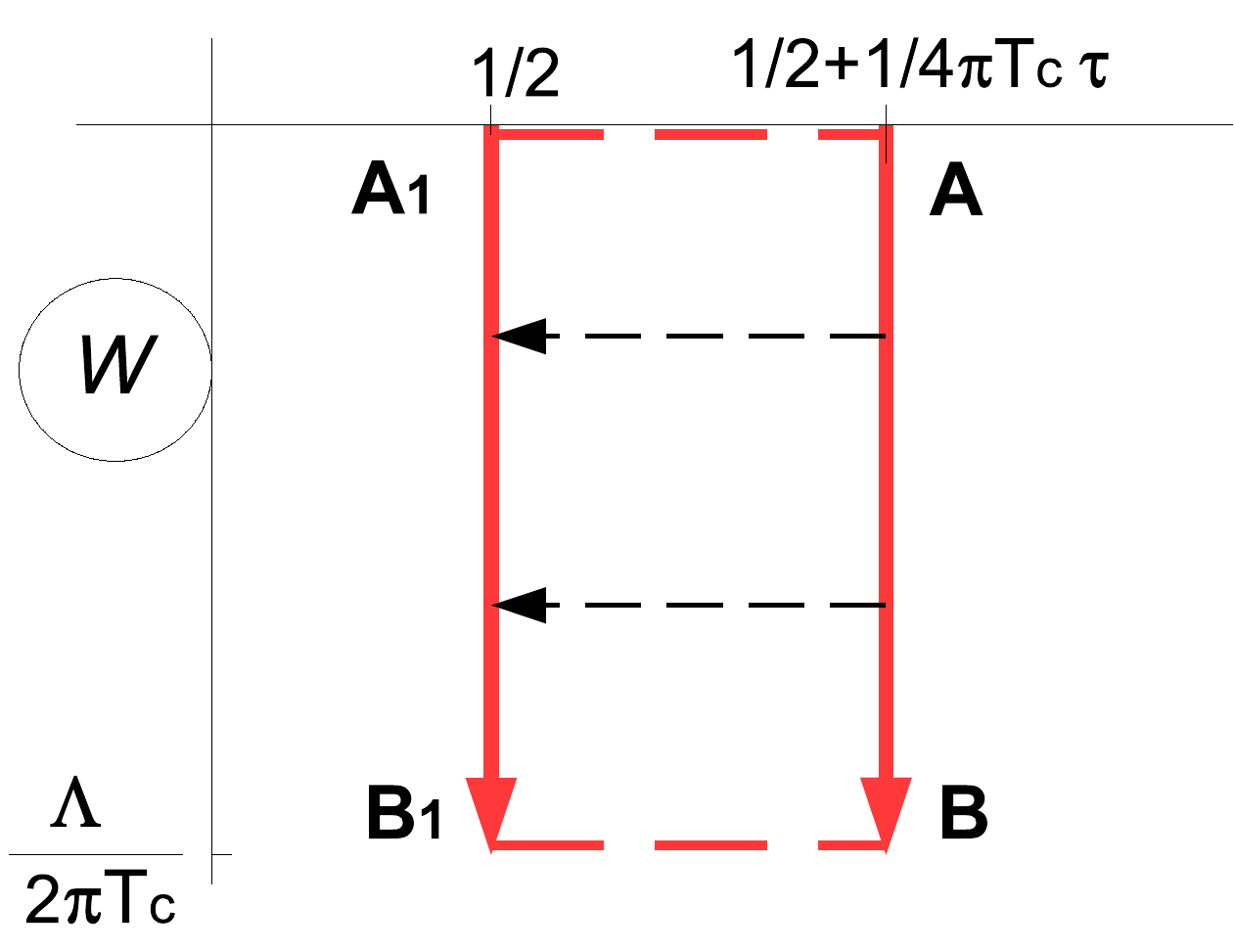}
  \caption{ \label{cp} Contours of integration in the complex plane $w$}
\end{figure}

In order to obtain Eq.(28) of the main text we will rely on equations (\ref{Kp}), (\ref{Tcc}) - (\ref{fnTc}), and use  the following relations for digamma function $\Psi(z)$:
\beq
\Psi\left(\frac{1}{2} + i x\right) -\Psi\left(\frac{1}{2} - i x\right) = i \pi \tanh (\pi x ),
\eeq
\beqa
\Psi\left(\frac{1}{2} + i x\right) +\Psi\left(\frac{1}{2} -i x\right) - \Psi\left(\frac{1}{2} \right) \nonumber\\
 =\left\{
\begin{array}{rcl}
16.8 x^2, &x \ll 1 &\\
2 \ln x - 2\Psi\left(\frac{1}{2}\right) -\frac{1}{12 x^2}, & x \gg 1 &.\\
\end{array}
\right.
\label{}
\eeqa
\beq
\int \left(\Psi\left(\frac{1}{2} + i x\right) +\Psi\left(\frac{1}{2} - ix\right)  \right) dx = i \ln \left( \frac{\Gamma(\frac{1}{2}-ix)}{\Gamma(\frac{1}{2}+ix)}\right),
\eeq
\beqa
\int_0^a \left(\Psi\left(\frac{1}{2} + i x\right) +\Psi\left(\frac{1}{2} - ix\right) - \Psi\left(\frac{1}{2} \right)\right) \frac{dx}{x} \nonumber \\
\backsimeq \ln^2 a - 2 \Psi\left(\frac{1}{2}\right) \ln a, \text{ for } a \gg 1,
\eeqa
We then rewrite Eq.~\ref{Tcc} in the form
\beqa
1&=& - \fint  \frac{d^2\kk}{(2\pi)^2}   \frac{(f(k, k_F))^2}{f(k_F, k_F)}\left[K_0(k) -\frac{1}{2(E_{k} - E_F)}\right] \nonumber \\
& &-\fint  \frac{d^2\kk}{(2\pi)^2}   \frac{(f(k, k_F))^2}{f(k_F, k_F)}\left[K(k)- K_0 (k)\right],
\eeqa
The first line in (A.5) is  the equation for the critical temperature $T_{c}^{0}$ in the absence of a disorder. The second  line contains $1/\tau$ corrections originating from the momentum dependance of the interaction
amplitude. Near the Fermi surface we have relations (\ref{nearTc}) and (\ref{fnTc}) and rewrite the first term in the second line of Eq.~(A.5) as
\begin{widetext}
\beq
\fint  \frac{d^2\kk}{(2\pi)^2}   \frac{(f(k, k_F))^2}{|f_0|}K(k)
 = \fint_0^{\Lambda} m\frac{d\xi}{2\pi}
\frac{1}{|f_0|}  \left[ \left(f_0 - \frac{8 d^2}{v_F} \xi\right)^2 -f_0^2\right] \left[ \frac{1}{2\pi}\frac{\Psi\left(\frac{1}{2} -\frac{i (\xi +\frac{i}{2\tau})}{2 \pi T_c}\right) - \Psi \left(\frac{1}{2}\right)}{-i \left(\xi + \frac{i}{2\tau}\right)}\right] + c.c.
\eeq
\end{widetext}
To calculate  the last integral  we use  analytical  properties of the $\Psi(w)$  in the complex plane
$w = Re w + i Im w$. This function is holomorphic  in the right semi-plane.  The  integration over
$d\xi$  can be considered as the integration along the  line AB  in Fig.~\ref{cp}, where
 $w = 1/2 + 1/ 4\pi T_c \tau - i \xi /2 \pi T_c$. As the integral along the closed contour $AB\rightarrow BB_1\rightarrow B_1A_1\rightarrow A_1A$ is equal to zero and the integrals along the lines
 $BB_1$ and $A_1A$ can be omitted \cite{noteA}, the integral along the line $AB$ is equal to the integral along the line $B_1A_1$. This is equivalent to the change $   w \to w - 1/4 \pi T_c \tau$, or $\xi \to \xi -i/2\tau$ in Eq.(A.6)
 which then takes the form

\begin{widetext}
\beq
 \fint_0^{\Lambda} m\frac{d\xi}{2\pi}
\frac{1}{|f_0|}  \left[ \left(f_0 - \frac{8 d^2}{v_F} \left(\xi - \frac{i}{2\tau}\right)  \right)^2 -f_0^2\right]\ \left[\frac{1}{2\pi} \frac{\Psi\left(\frac{1}{2} -\frac{i \xi }{2 \pi T_c}\right) - \Psi \left(\frac{1}{2}\right)}{-i \xi }\right]+ c.c.,
\eeq
where we put an upper bound $\Lambda\sim \mu$ having in mind that the integral will be logarithmically divergent.
\end{widetext}

Substituting the integral of Eq.(A.7) and its complex conjugated value into (A.5) we obtain
 \beqa
 1-&   \fint_0^{\Lambda} m\frac{d\xi}{2\pi}  \frac{(f(k, k_F))^2}{|f_0|}\left[K_0(k) -\frac{1}{2\xi}\right] = \nonumber \\
+& \fint_0^{\Lambda} m\frac{d\xi}{2\pi}\frac{1}{\tau}\frac{8 d_0^2}{v_f}\frac{\Psi\left(\frac{1}{2} +\frac{i \xi }{2 \pi T_c}\right) +\Psi\left(\frac{1}{2} -\frac{i \xi }{2 \pi T_c}\right) -2 \Psi\left(\frac{1}{2}\right)}{2\pi \xi} \nonumber\\
+& \fint_0^{\Lambda} m\frac{d\xi}{2\pi}\frac{1}{\tau}\frac{64 d_0^4}{|f_0| v_f^2}\frac{\Psi\left(\frac{1}{2} +\frac{i \xi }{2 \pi T_c}\right) +\Psi\left(\frac{1}{2} -\frac{i \xi }{2 \pi T_c}\right) -2 \Psi\left(\frac{1}{2}\right)}{2\pi } \nonumber\\
&- \fint_0^{\Lambda} m\frac{d\xi}{2\pi} \frac{1}{\tau^2} \frac{8d_0^4}{|f_0| v_F^2} \frac{\tanh\frac{\xi}{2 T_c}}{\xi}.
\eeqa

The first line of Eq.(A.5) gives a standard expression   $\lambda \ln (T_c /T_C^0)$, where $T_c^0=\frac{2 \mu e^{C}}{\pi} \exp (-1/\lambda)$ is the critical temperature in the absence of disorder.
The second line is calculated using Eq(A.4) and it gives :
\beq
 \frac{1}{k_F l} \frac{2 k_F r_*}{\pi^2}\left[ \ln^2\frac{\mu}{2\pi T_c} -2\Psi\left(\frac{1}{2}\right) \ln\frac{\mu}{2\pi T_c} \right].
 \eeq
 The third line is calculated  by the use of Eq.(A.3):
\beq
\frac{1}{k_F l} \frac{8 (k_F r_*)^2}{\pi^3 \lambda} \frac{ \mu}{v_F p_F} \left[2\ln\frac{\mu}{2\pi T_c} -2-2\Psi\left(\frac{1}{2}\right)\right].
\eeq
The  forth line contains a standard integral and  it gives
\beq
-\frac{1}{(k_Fl)^2}  \frac{(k_F r_*)^2}{\pi^2 \lambda}  \ln  \frac{2 \mu e^{C}}{\pi T_c}.
\eeq
Substituting relations (A.9)-(A.11) into Eq.(A.8) we obtain Eq.~(\ref{eqnTc}) of the main text.

Note that we used approximate relations (\ref{nearTc}) and (\ref{fnTc}) valid near the Fermi surface. However, we checked that the use of exact expressions for $\xi$ and $F(k,k_F)$ leads to practically the same results.


\begin{thebibliography}{99}
\bibitem{Note} The quantity $\Gamma/\gamma$ is given by the series $1 + \gamma B_0(\varepsilon_n) + [\gamma B_0(\varepsilon_n)]^2 + ... $,  where $B_0(\varepsilon_n)$ is determined by Eq.(\ref{GG}).
\bibitem{noteA} The integral along the line $BB_1$ is much smaller than the integral along the line $A_1A$. The ratio of the latter to the main term of Eq.(\ref{eqnTc}) is smaller than
$\ln^2(1/4\pi T_c\tau)/\ln^2(E_F/2\pi T_c)$. For all realistic parameters $\lambda$ and $k_Fl$ this ratio is much smaller than unity. Moreover, the integral along the line $A_1A$ is also
smaller than (or of the order of) the other terms of Eq.(\ref{eqnTc}).
\end{thebibliography}
  \end{document}